# The Effect of Chemical Doping and Hydrostatic Pressure on $T_c$ of $Y_{1-y}Ca_yBa_2Cu_3O_x$ Single Crystals

S.I. Schlachter[ab]*, W.H. Fietz[a], K. Grube[a], Th. Wolf[a], B. Obst[a], P. Schweiss[c], M. Kläser[c]

[a] *Forschungszentrum Karlsruhe, Institut für Technische Physik, Postfach 3640, 76021 Karlsruhe, Germany*
[b] *Universität Karlsruhe, Fakultät für Physik, 76128 Karlsruhe, Germany*
[c] *Forschungszentrum Karlsruhe, Institut für Festkörperphysik, Postfach 3640, 76021 Karlsruhe, Germany*

**ABSTRACT**

We performed susceptibility measurements on $Y_{1-y}Ca_yBa_2Cu_3O_x$ single crystals under high He pressure. For each Ca content various samples with different oxygen contents have been prepared to probe the influence of Ca on $T_c(x)$, $dT_c/dp(x)$ and $T_{c,max}$. Starting from the parabolic $T_c(n_h)$ behavior we calculated $n_h$ values from $T_c$ and $T_{c,max}$ for each sample. It is shown that in the overdoped region $dT_c/dp$ can be described by a pressure induced charge transfer with $dn_h/dp \approx 3.7 \cdot 10^{-3}$ GPa$^{-1}$ and a $dT_{c,max}/dp$ value of 0.8 K/GPa, irrespective of the Ca content.

In the underdoped region additional pressure effects lead to a peak in $dT_c/dp$ at $\approx 0.11$ holes/CuO$_2$ plane. However, with increasing Ca content this peak is strongly depressed. This is explained in terms of an increasing disorder in the CuO chain system due to doping. Deviations in $dT_c/dp$ at very low $n_h$ values can be assigned to the ortho II ordering in the CuO chain system.

*PACS:* 74.62.-c, 74.62.Fj, 74.62.Dh, 74.72.BK.
*Keywords:* High-$T_c$ superconductor; Charge transfer; Doping; Oxygen order; High pressure

## 1. Introduction

Since the discovery of $YBa_2Cu_3O_x$ (Y123) in 1987 [1] having a transition temperature to superconductivity, $T_c$, above liquid nitrogen temperature, efforts have been made to understand the mechanism of high-temperature superconductivity. Many new high-$T_c$ compounds have been found since those days, most of them showing a characteristic layered structure with one or more CuO$_2$ planes in which superconductivity takes place. A distinctive feature of the high-$T_c$ superconductors is the strong dependence of the antiferromagnetic and the superconducting properties on the charge carrier concentration, $n_h$, (holes per CuO$_2$ plane). For very low $n_h$ values a long-range antiferromagnetic order is observed. Increasing $n_h$ leads to a destruction of this antiferromagnetic order. As $n_h$ is raised further superconductivity appears, with $T_c$ increasing from zero to a maximum transition temperature, $T_{c,max}$, and decreasing again in the so-called overdoped region. There were several suppositions that $T_c(n_h)$ has an inverted parabolic shape [2-9]. Presland et al. [10] and Tallon et al. [11] reported this parabolic shape to be a universal relation if the reduced temperature $T_c/T_{c,max}$ is plotted versus $n_h$:

$$\frac{T_c}{T_{c,max}} = 1 - \left(\frac{n_h - n_{h,opt}}{0.11}\right)^2 \quad (1)$$

* Corresponding author. Forschungszentrum Karlsruhe, Institut für Technische Physik, Postfach 3640, 76021 Karlsruhe, Germany; Tel.: +49-7247-82-3554; fax +49-7247-82-2849.
E-mail address: sonja.schlachter@itp.fzk.de (S.I. Schlachter)

The maximum transition temperature $T_{c,max}$ at optimum doping $n_{h,opt} \approx 0.16$ is different for each compound, but a systematics for $T_{c,max}$ may be found by varying the chemical composition. For example in the system $RBa_2Cu_3O_x$ with R = Yb, Er, Dy, Y, Gd, Eu and Nd the achievable transition temperature $T_c$ increases with increasing R-ion radius [12-14] and by that also $T_{c,max}$. To determine the particular $T_{c,max}$ value extensive experiments on samples with different hole concentrations in the $CuO_2$ planes are necessary. In review articles [15-17] a lot of doping experiments are reported.

Apart from chemical doping the hole concentration $n_h$ can be varied by the application of high pressure, as has been suggested by Mori et al. [18] for Tl-Ba-Ca-Cu-O and by van Eenige et al. [19] and Jorgensen et al. [20] for $YBa_2Cu_3O_x$. The latter system offers the possibility to vary $n_h$ from zero to the slightly overdoped region by changing the oxygen content $x$ from 6 to 7. Under pressure the apical oxygen atom located between the CuO chain and $CuO_2$ plane is shifted towards the plane leading to a charge transfer from the chain to the plane at a constant chemical composition [20]. Hence, the chains may be viewed as a charge reservoir. Hall measurements [21-27] confirm the enhancement of the hole concentration in the planes by the application of pressure. A lot of calculations based on experimental observations and theoretical models show that pressure dopes the $CuO_2$ planes from a charge reservoir. Data for the pressure induced charge transfer are given for example in [28-43]. Murayama et al. [21] showed from Hall coefficient measurements on Y124, Bi2212 and Tl221 that $T_c$ follows the $T_c(n_h)$ parabola via pressure induced doping, whereas for doped La214 a complete shift of the parabola to higher $T_c$ values and no pressure induced doping effect was found [44]. Almasan et al. [35] and Iye [45] extended the simple charge transfer model assuming that the effect of pressure on $T_c$ consists of a change of the carrier density and an additional shift of the complete $T_c$ parabola to higher $T_c$ values expressed by $dT_{c,max}/dp$. The influence of pressure on $T_{c,max}$ can easily be determined when $T_{c,max}$ is known at ambient pressure because in the maximum of the parabola the $T_c$ variation due to a pressure induced change of the hole concentration is zero. The pressure dependence of this $T_{c,max}$ value was found to be positive for various high-$T_c$ superconductors [37, 41-43, 46-53]. More information about $dT_{c,max}/dp$ and $n_h(p)$ may be found in the review articles [15, 16]. In addition we will give some values for $dn_h/dp$ and $dT_{c,max}/dp$ in the discussion of our results.

With increasing knowledge it is getting more and more questionable, whether the effect of pressure on the superconducting transition temperature, $dT_c/dp$, in different high-$T_c$ systems can be described in the same manner. For most high-$T_c$ materials the uniaxial pressure effect in $a/b$- and in $c$-direction is anisotropic [54-58] and in Y123 the effect of pressure in $a$- and $b$-direction is even of opposite sign [59-63] due to the orthorhombic structure caused by the CuO chains. Such a delicate effect has naturally not been found in any tetragonal high-$T_c$ material and even the very similar $YBa_2Cu_4O_8$ which has a doubled CuO chain with a fixed oxygen content does not show pressure effects in $a$- and $b$-axis directions with opposite signs [54]. In this work we therefore concentrate on the effect of pressure in the doped Y123 system that is well known and offers repro-ducible good single crystals. The unique CuO chains with variable oxygen content give the possibility to change $n_h$ from zero to the slightly overdoped region with a single doping channel, and by substituting $Ca^{2+}$ for $Y^{3+}$ even the heavily overdoped region is accessible.

A lot of experiments have been performed to determine $dT_c/dp$ of doped Y123 compounds [19, 21, 35, 37, 43, 47, 49, 58, 64-71]. In the beginning of these investigations the results from different groups showed large scatter. Early results from electron diffraction [72] showed the appearance of oxygen ordering in the CuO chains, which was investigated in detail via quench experiments [73, 74]. An influence of pressure on oxygen ordering was found for the Tl221 compound at temperatures even below 40 K [75-79]. Similar processes were seen for Y123 under pressure [69, 80-85] but in contrast to Tl221 it was shown by $T_c(p)$ determinations that the oxygen ordering effect was absent at temperatures below 240 K [80]. As those authors had to go above $T_c$ to determine the appropriate $T_c$ values of their samples they mentioned possible low temperature relaxation processes that might have caused unknown structural changes. Klehe et al. [78] argued that low temperature oxygen ordering may exist and that it could be responsible for the peak in $dT_c/dp$ at an oxygen content of $x$=6.7 [82]. However, pressure induced oxygen ordering was absent for an Y123 sample with an extremely low oxygen content x=6.41, as shown by Sadewasser et al. [79]. Due to the low $T_c$ of that sample it was proofed that no oxygen ordering effects occur between temperatures of 14 K and 250 K. Although relaxation processes at temperatures below 14 K can not be ruled out, it is clear that experiments on oxygen deficient Y123 with pressure application above or below 240 K will result in different $dT_c/dp$ values due to the resulting enhanced $T_c$ values from oxygen ordering [81, 82]. This is one reason for the large scatter in the comparison of early experiments.

Most results for $dT_c/dp$ on doped Y123 were plotted as a function of chemical doping because of problems in determining $n_h$. Miyatake et al. [68] reported $dT_c/dp(n_h)$ values for the $(Yb_{0.7}Ca_{0.3})(Ba_{0.8}Sr_{0.2})_2Cu_3O_x$ system, estimating $n_h$ from the doping by empirical relations. They assigned 1/3 of the charge carriers to each of the 2 planes and the CuO chain and obtained a nonlinear $dT_c/dp(n_h)$. However, Miyatake et al. presumably applied pressure at room temperature, which includes oxygen ordering, too.

Neumeier and Zimmermann [37] avoided the problem of oxygen ordering by using almost fully oxygenated samples, only. Using $Y_{1-y}Ca_yBa_{2-z}La_zCu_3O_x$ they made the under- and overdoped range accessible by substituting $La^{3+}$ for $Ba^{2+}$ and $Ca^{2+}$ for $Y^{3+}$ and they assumed to create or destroy one hole per unit cell by doping with Ca or La, respectively. They showed $dT_c/dp$ to be linear as a function of doping, resulting in a charge transfer of approximately 0.011 holes per GPa and unit cell, and a $dT_{c,max}/dp = 0.96$ K/GPa. These valuable results, however, have to be reexamined in the light of the information that is available today. For instance it is now known that doping with Ca does reduce the $T_{c,max}$ value [8, 11, 49] which implies that with increasing Ca-content the $T_c$ value is not only determined by the increasing hole doping, but is influenced by the $T_c$ suppression from Ca, too. In addition with fixed annealing conditions the resulting oxygen content $x$ of a $Y_{1-y}Ca_yBa_2Cu_3O_x$ is reduced [86] by $\approx y/2$, and for larger Ca contents Ca does no longer tend to dope the Y-site only [87-90]. The orthorhombicity of $Y_{1-y}Ca_yBa_2Cu_3O_x$ is influenced by Ca, too, because with increasing Ca content oxygen atoms in the vicinity of a Ca atom tend to occupy not only O(1) sites (in chain) but also O(5) sites (out of chain) [88, 91]. As a consequence the disorder in the CuO chain sublattice is increased with increasing Ca content, making it more difficult to achieve a well ordered CuO chain system in the *ortho* I or the *ortho* II phase. A $T_c$ decrease may also be the consequence of an Al contamination [92-94] of the Ca doped samples of Neumeier and Zimmermann, because this is unavoidable using $Al_2O_3$ crucibles for sample preparation [95].

Hikita and Ogasawara [86] presented investigations of $T_c$ and $dT_c/dp$ on Ca- and La-doped Y123. For Ca doping they demonstrated the problems of an additive oxygen deficiency by introducing Ca in Y123. With a small substitution of 0.25% La for Ba in fully oxygenated and thus overdoped Y123 they obtained only a slight $T_c$ increase to $T_{c,max}$=90.7 K. A further increase of the La content to 15% reduced $T_c$ to 83.3 K. A similar result was reported by Tallon and Flower [8] with a $T_{c,max}$ reduction from 91.8 K for the optimally doped sample $YBa_2Cu_3O_x$ to 89 K for $Y(Ba_{0.9}La_{0.1})_2Cu_3O_7$ and 82 K for $Y(Ba_{0.8}La_{0.2})_2Cu_3O_7$.

In contrast to these results Neumeier and Zimmermann [37] observed $T_c$ of 94.2 K for an Y123 sample doped with 7% La. This value is significantly higher than the $T_{c,max}$ for optimal Y123 doped by oxygen, only [96, 97]. One possible explanation of this discrepancy may be an enhancement of the $T_c$ values of the La doped samples due to a larger effective ion radius on the Y site caused by La substitution [12-14], but if that would be true the nominal doping has to be corrected for these La atoms.

The influence of the Ca or La doping on the oxygen content or the uncertainty of the true position of the doped atoms demonstrate the problems of estimating the hole density in the $CuO_2$ plane from the nominal chemical doping. In addition the destruction or generation of holes by La or Ca doping may not result in similar changes $|\Delta n_h|$ of the hole concentration within the $CuO_2$ planes, because holes from Ca doping are found to enter the planes only, whereas holes created by other doping channels are distributed over the two $CuO_2$ planes, the two apical-oxygen atoms and the CuO chain [98, 99].

Another question arises from the linearity that is found for $dT_c/dp$ as a function of doping in the investigation of Neumeier and Zimmermann [37]. Several experiments on Y123 have shown that $dT_c/dp$ is nonlinear with decreasing oxygen content for $6.75<x<7.0$ [35, 61, 69, 71, 81], although at least for oxygen contents from $x=6.8$ to 6.95 the change in the oxygen content should be nearly proportional to the $n_h$ change. In this range each added oxygen atom should be set to the end of a Cu-O-chain which is a consequence of oxygen ordering. Thus each added oxygen should change one $Cu^{1+}$ to a $Cu^{2+}$ and in addition create one hole in the unit cell. This nonlinearity has also been found for Ca-doped Y123 and is assigned to the uniaxial pressure effect in a/b-axis direction, whereas c-axis pressure is responsible for charge transfer [43, 49, 58]. However, the existence of the nonlinearity may be coupled to the choice of the doping channel.

To clarify these open questions we have performed investigations on $Y_{1-y}Ca_yBa_2Cu_3O_x$ samples with Ca contents from 0% to 22%. To avoid effects of segregation at grain boundaries we performed experiments on single crystals, only. We measured $T_c$ and $dT_c/dp$ at numerous oxygen contents for each Ca content separately to study the influence of a single doping channel. From these results we obtained the valid $T_{c,max}$ value for each Ca content. To demonstrate the influence of Al contamination we included a sample grown in an $Al_2O_3$ crucible, too.

## 2. Experimental details

$Y_{1-y}Ca_yBa_2Cu_3O_x$ single crystals were synthesized in Y or Ca stabilized $ZrO_2$ crucibles using $Y_2O_3$, BaO ($BaCO_3$), CuO and CaO ($CaCO_3$) as starting materials. Crystal growth took place during slow cooling of the melt from 1020°C to approximately 940°C in atmospheres with appropriate oxygen partial pressures to achieve that Ca dopes the Y site only. After decanting the melt within the furnace the crystals were cooled to room temperature. The Ca contents, $y$, of the $Y_{1-y}Ca_yBa_2Cu_3O_x$ single crystals were 0.00, 0.04, 0.08, 0.11, 0.20 and 0.22. Under the assumption of a perfect 123 structure, quantitative EDX analysis with an accuracy of $\leq 1\%$ showed that in the samples with low Ca contents ($y \leq 0.11$) the $Ca^{2+}$ ions occupy only $Y^{3+}$ sites whereas in the samples with higher Ca contents ($y \geq 0.20$) up to 2% of the $Ca^{2+}$ ions are found on $Ba^{2+}$ positions, too. These results are in accordance with additional neutron diffraction measurements.

Different oxygen contents $x$ were adjusted by annealing the samples in flowing oxygen or an appropriate $N_2/O_2$ mixture at various temperatures and oxygen partial pressures, $P_{O_2}$. At the end of the annealing process the samples were quenched in liquid nitrogen and then stored at room temperature for at least one week to adjust an equilibrium in oxygen ordering. Because of the small size of the single crystals we could not determine the oxygen content $x$ by weighing or by chemical methods. Therefore, $x$ was estimated from the annealing conditions, using Ca content dependent isotherm curves $\log P_{O_2}$ vs. oxygen content [87, 88]. The absolute oxygen contents obtained from these isotherms have been checked for selected samples via neutron scattering and are reliable enough to show trends in $T_c(x)$ and $dT_c/dp(x)$.

Measurements of $T_c$ and $dT_c/dp$ were performed in a CuBe pressure cell using He gas as pressure transmitting medium. In this way true hydrostatic pressure conditions up to 0.6 GPa could be achieved. $T_c$ was determined by ac susceptibility measurements using the peak in $\chi''$ as $T_c$ criterion. In order to avoid pressure induced oxygen ordering effects [79-83] pressure was applied at low temperature ($T \leq 110$ K) and the sample was kept at temperatures below 110 K during the whole experiment. At the beginning and the end of the experiment $T_c$ was measured at ambient pressure. A comparison of these $T_c$ values never gave any hints of pressure induced oxygen ordering effects or damages of the sample due to pressure.

## 3. Results and discussion

### 3.1. Chemical doping

In Fig. 1 the obtained $T_c$ values of $Y_{1-y}Ca_yBa_2Cu_3O_x$ single crystals with $y$ = 0.00, 0.04, 0.08, 0.11, 0.20 and 0.22 are plotted vs. oxygen content $x$. With increasing Ca content the oxygen content $x$ at $T_{c,max}$ is shifted towards lower $x$ values as shown in Fig. 2a. The shift is expected because by substituting $Ca^{2+}$ for $Y^{3+}$ additional holes are introduced to the system. Therefore, optimum doping $n_{h,opt}$ can already be achieved at lower oxygen contents.

A further increase of $x$ allows a deep penetration into the overdoped region, decreasing $T_c$ thereby more and more. $T_c$ of a sample with $y$ = 0.22, for example, dropped from 82.1 K at optimum doping to 61.9 K at an oxygen content of $x \approx 6.92$.

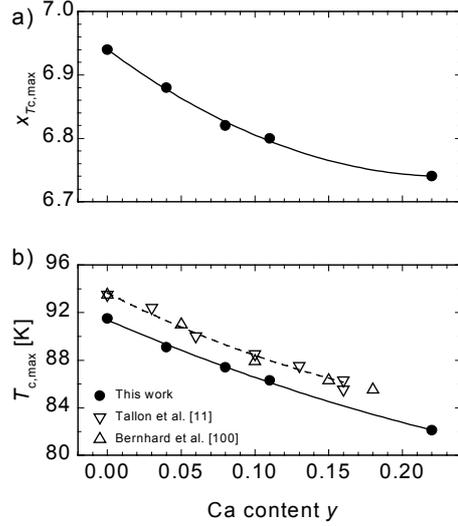

Fig. 2. (a) Oxygen content $x_{Tc,max}$ at maximum transition temperature vs. Ca content. (b) The maximum transition temperature $T_{c,max}$ of $Y_{1-y}Ca_yBa_2Cu_3O_x$ at ambient pressure vs. Ca content. The open symbols show data from Tallon et al. [11] and Bernhard and Tallon [100] corrected to the effective Ca contents, i.e. Ca on Y sites, given by these authors.

The decrease of $T_{c,max}$ with increasing Ca content (Fig. 2b) has already been reported by Tallon et al. [11] and Bernhard and Tallon [100]. A $T_{c,max}$ offset between those values from polycrystalline samples and our single crystal data may have its origin either in the different sample material or in the use of different $T_c$ criteria, however, the $T_c$ decrease with Ca doping is identical in all three experiments. The decrease of $T_{c,max}$ can not be ascribed to the ion size effect [12-14] that determines $T_{c,max}$ in $RBa_2Cu_3O_x$, because from the bigger size of $Ca^{2+}$ compared to $Y^{3+}$ an increase of $T_{c,max}$ with increasing Ca content would be expected. On the basis of NEXAFS measurements Merz et al. [99] explained the $T_{c,max}$ decrease by a reduced oxygen content in the CuO chains of the optimally doped $Y_{1-y}Ca_yBa_2Cu_3O_x$, which may decrease the coupling of $CuO_2$ planes via the Apex - CuO chain - Apex system.

### 3.2. Pressure induced hole doping of the $CuO_2$ planes

When pressure is applied to a HTSC the resulting change of the superconducting transition temperature depends strongly on the structure, the chemical composition and the hole concentration which is influenced by the chemical composition and the structure, too. Fig. 3 shows the influence of pressure on $T_c$ of $Y_{1-y}Ca_yBa_2Cu_3O_x$, i.e. $dT_c/dp$ vs. the oxygen content $x$. With decreasing oxygen content, $dT_c/dp$ of the Ca free samples increases from small negative values at high oxygen contents to large positive values of approximately 7 K/GPa. At about $x = 6.7$, $dT_c/dp$ drops to smaller values of approximately 3 K/GPa. Increasing the Ca content of the samples decreases the maximum in $dT_c/dp$, flattens the peak and moves it from $x = 6.7$ for the Ca-free samples to slightly lower oxygen contents for samples with higher Ca contents. For $y = 0.22$ only a plateau-like behavior below $x = 6.63$ is observed. With increasing Ca content the zero crossing of $dT_c/dp$ is shifted from $x \approx 6.95$ for the Ca-free samples to $x \approx 6.82$ for the samples with $y = 0.22$.

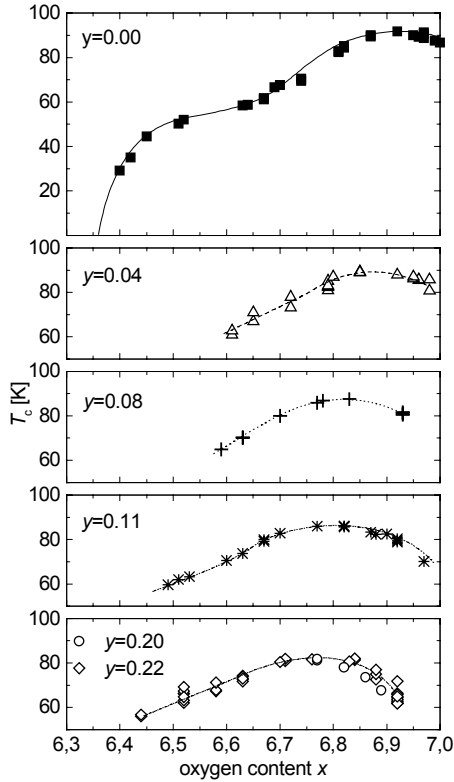

Fig. 1. $T_c$ of $Y_{1-y}Ca_yBa_2Cu_3O_x$ vs. oxygen content at ambient pressure for different Ca contents $y$ = 0.00, 0.04, 0.08, 0.11, 0.20 and 0.22.

According to Eq. (1) $T_c$ can be described by (1) the hole concentration, $n_h$, (2) the maximum transition temperature, $T_{c,max}$, and (3) the optimum hole concentration, $n_{h,opt}$, that is $T_c=T_c(n_h, T_{c,max}, n_{h,opt})$. If we assume $n_h$, $T_{c,max}$ and $n_{h,opt}$ to depend on pressure, the application of pressure then changes $T_c$ in the following way:

$$\frac{dT_c}{dp} = \left(\frac{\partial T_c}{\partial n_h}\right)_{T_{c,max}, n_{h,opt}} \cdot \frac{dn_h}{dp} + \left(\frac{\partial T_c}{\partial T_{c,max}}\right)_{n_h, n_{h,opt}}$$

$$\cdot \frac{dT_{c,max}}{dp} + \left(\frac{\partial T_c}{\partial n_{h,opt}}\right)_{n_h, T_{c,max}} \cdot \frac{dn_{h,opt}}{dp} \quad (2)$$

According to Eq. (2) pressure induced variations of $n_h$, $n_{h,opt}$ and $T_{c,max}$ effect $T_c$ in the following way:

1. the positive $dn_h/dp$ leads to changes of $T_c$ that follow the $T_c(n_h)$ parabola, i.e. pressure induced charge transfer increases $T_c$ in the underdoped region ($n_h < n_{h,opt}$) and decreases $T_c$ in the overdoped region ($n_h > n_{h,opt}$). For optimally doped samples a small variation $\delta n_h$ does not affect $T_c$ at all.
2. $dT_{c,max}/dp$ moves the maximum transition temperature $T_{c,max}$ to higher or lower values. As $dT_{c,max}/dp$ is usually found to be positive, this effect adds to the $T_c$ increase from item (1) in the underdoped region but subtracts in the overdoped region.
3. $dn_{h,opt}/dp$ moves the whole parabola to higher or lower hole concentrations.

Differentiating explicitly Eq. (1) with respect to pressure we find for $dT_c/dp$:

$$\frac{dT_c}{dp} = \left(\frac{T_c}{T_{c,max}}\right) \cdot \left(\frac{dT_{c,max}}{dp}\right) - T_{c,max}$$

$$\cdot \left(\frac{n_h - n_{h,opt}}{6.05 \cdot 10^{-3}}\right) \cdot \frac{d(n_h - n_{h,opt})}{dp} \quad (3)$$

Following Goldschmidt et al. [41] we multiply Eq. (3) with $T_{c,max}/T_c$. For constant values of $dn_h/dp$, $dT_{c,max}/dp$ and $dn_{h,opt}/dp$, in a plot of $(T_{c,max}/T_c) \cdot (dT_c/dp)$ versus $(T_{c,max}^2/T_c) \cdot (n_h - n_{h,opt})/(6.05 \cdot 10^{-3})$ we should find a straight line $y = m \cdot x + b$ with $m = d(n_h - n_{h,opt})/dp$ and $b = dT_{c,max}/dp$. For our samples the assumption of constant $dn_h/dp$, $dT_{c,max}/dp$ and/or $dn_{h,opt}/dp$ may be questionable, however, if they were not the same for all samples and doping states we would not expect a linear behavior.

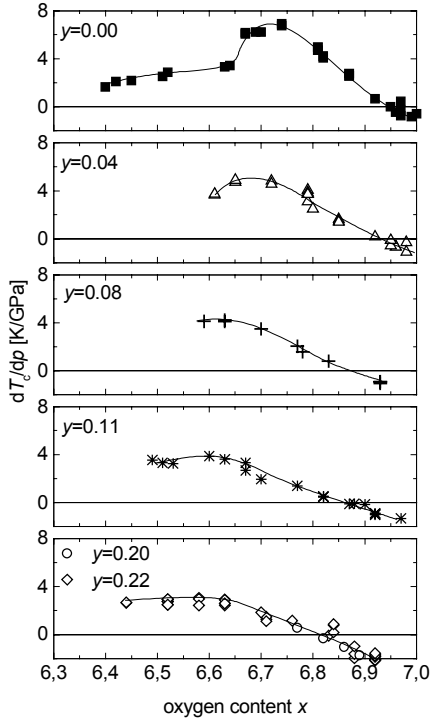

Fig. 3. $dT_c/dp$ of the $Y_{1-y}Ca_yBa_2Cu_3O_x$ samples versus oxygen content.

To perform this special plot we need information about the hole concentration in the $CuO_2$ planes. Due to the influence of Ca on the final oxygen content when samples are annealed, the problem of estimating the hole concentration in the $CuO_2$ planes from chemical doping as mentioned in the introduction and the principal problems of an accurate determination when small single crystals are used we did not try to estimate the hole concentration from the nominal chemical doping. Following Obertelli et al. [101] we made use of Eq. (1) and the measured $T_c$ and $T_{c,max}$ values shown in Fig. 1 resulting in:

$$n_h = n_{h,opt} \pm 0.11 \cdot \sqrt{1 - T_c/T_{c,max}} \qquad (4)$$

with $n_{h,opt} = 0.16$ for HTSCs. In Eq. (4) the negative sign has to be taken for underdoped samples ($n_h < n_{h,opt}$) and the positive sign for $n_h > n_{h,opt}$. Whether a sample is under- or overdoped can be directly inferred from the $T_c(x)$ curves in Fig. 1 where the oxygen content of the samples was estimated from the annealing conditions.

With this information we are now able to plot our data in the above mentioned way resulting in Fig. 4a: In the optimally doped and overdoped region we indeed find a linear behavior for the samples with various Ca contents confirming the assumption of constant values for $dn_h/dp$, $dn_{h,opt}/dp$ and $dT_{c,max}/dp$. Therefore, in this region we can easily extract the values for $d(n_h-n_{h,opt})/dp$ and $dT_{c,max}/dp$ from the linear fit to the data of the optimally doped and overdoped samples (straight line in Fig. 4a). According to Eq. (1) the optimum hole concentration for various HTSCs is $n_{h,opt} = 0.16$, despite the variation in structure of the different compounds. For this reason it can be assumed that in $Y_{1-y}Ca_yBa_2Cu_3O_x$ the comparatively small structural changes due to pressure do not alter $n_{h,opt}$ massively, i.e. $dn_{h,opt}/dp \approx 0$. Therefore, for optimally doped and overdoped $Y_{1-y}Ca_yBa_2Cu_3O_x$ single crystals with Ca contents ranging from zero to 22% we find a uniform value for the pressure induced charge transfer to the $CuO_2$ plane of

$$\frac{dn_h}{dp} \approx 3.7 \cdot 10^{-3}\,GPa^{-1} \qquad (5)$$

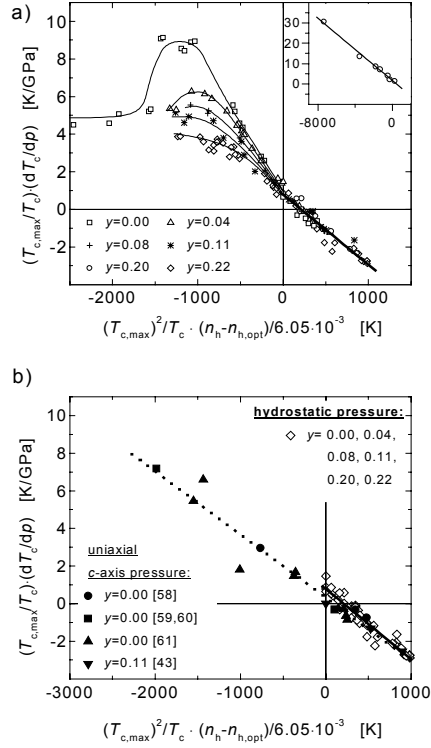

Fig. 4. (a) Special plot of our $Y_{1-y}Ca_yBa_2Cu_3O_x$ data, deduced from the charge transfer model for extracting $d(n_h-n_{h,opt})/dp$ and $dT_{c,max}/dp$. The inset shows corresponding values for $(Ca_zLa_{1-z})(Ba_{1.75-z}La_{0.25+z})Cu_3O_x$, calculated from the data of Goldschmidt et al. [41] with the straight line as a fit to those values. (b) Uniaxial $c$-axis pressure data of $YBa_2Cu_3O_x$ [43, 58-61] and $Y_{0.89}Ca_{0.11}Ba_2Cu_3O_x$ [43] and hydrostatic pressure data of over- and optimally doped $Y_{1-y}Ca_yBa_2Cu_3O_x$ plotted in the same way as in a) to show that the hydrostatic pressure effect in the overdoped region and the $c$-axis pressure effect $dT_c/dp_c$ are dominated by pressure induced charge transfer. The dashed line and the solid line are linear fits to the uniaxial and to the hydrostatic pressure data, respectively.

This value for the pressure induced hole transfer is in good agreement with data from the literature (e.g $6 \cdot 10^{-3}\,GPa^{-1}$ [61], $5 \cdot 10^{-3}\,GPa^{-1}$ [43] and $4 \cdot 10^{-3}\,GPa^{-1}$ [34]).

From the plot in Fig. 4a the variation of the maximum transition temperature for these various samples was found to be

$$\frac{dT_{c,max}}{dp} \approx 0.8\,K/GPa \qquad (6)$$

which is in good agreement with literature data for doped Y123, too (e.g. 1 K/GPa [86], 1 K/GPa [43], 0.89 K/GPa [47], 0.96 K/GPa [37], 0.4 K/GPa [61]).

In the underdoped region, however, the curves in Fig. 4 do not show a unique linear behavior but split into separate lines. Especially the samples with low Ca content show a strong deviation from the linear behavior. For these samples a pronounced maximum can be seen in Fig. 4a. With increasing Ca content the peak flattens out.

To show values from an almost isostructural compound we used data for the charge balanced material $(Ca_zLa_{1-z})(Ba_{1.75-z}La_{0.25+z})Cu_3O_x$ (CaLaBaCuO) from Goldschmidt et al. [41]. They needed an additional charge by oxygen doping of $11\cdot10^{-3}$ $|e|$ to balance the charge transfer per GPa. To show these data in Fig. 4a we calculated $n_h$ for those samples from $T_c$ and $T_{c,max}$ as mentioned above. The obtained data points cover a much wider range than our data due to the low $T_c$ values of the two most underdoped samples. Therefore, we have plotted these values in the inset of Fig. 4a and obtain a linear dependence with no indication for a peak in $dT_c/dp$. In comparison to our samples the linear dependence gives an enhanced $dT_{c,max}/dp$ of 1.7 K/GPa and an almost identical charge transfer of $3.8\cdot10^{-3}$ GPa$^{-1}$. The main structural difference of this compound in comparison with YBa$_2$Cu$_3$O$_x$ is the almost complete disorder in the CuO chains which is expressed by the tetragonal structure and the possibility to go beyond an oxygen content of $x = 7$. Therefore, it seems reasonable that the peak for the Y123 samples is suppressed with increasing Ca-content by the increasing disorder in the CuO chain subsystem.

For a better understanding of the origin of the pressure effect in general it is useful to look at the uniaxial pressure effects $dT_c/dp_i$ from which the hydrostatic pressure effect $dT_c/dp$ is composed:

$$\frac{dT_c}{dp} = \sum_i \frac{dT_c}{dp_i} \qquad (i = a,b,c) \qquad (7)$$

Uniaxial pressure effects of YBa$_2$Cu$_3$O$_x$ were measured directly by Ludwig et al. [58] and Welp et al. [59, 60]. Uniaxial pressure effects of YBa$_2$Cu$_3$O$_x$ samples with $6.5 < x < 7.0$ could also be deduced from measurements of the thermal expansion using Ehrenfests relation [61-63]. From these determinations of $dT_c/dp_i$ it can be deduced that the uniaxial pressure effect in c-axis direction, $dT_c/dp_c$, decreases almost linearly with increasing oxygen content of the samples with a zero crossing for almost optimally doped YBa$_2$Cu$_3$O$_x$ samples ($x \approx 6.9$). This is exactly what would be expected from pure charge-transfer behavior and is consistent with the fact that compression of the soft c-axis causes strong variations of bond lengths leading to an effective charge transfer from the CuO chains to the CuO$_2$ planes [20, 21, 102]. In Fig. 4b we have plotted the uniaxial c-axis pressure data of YBa$_2$Cu$_3$O$_x$ that were obtained from Ref. [58-61] and the uniaxial c-axis pressure data of Y$_{0.89}$Ca$_{0.11}$Ba$_2$Cu$_3$O$_x$ from Ref. [43] as well as the hydrostatic pressure data of our overdoped Y$_{1-y}$Ca$_y$Ba$_2$Cu$_3$O$_x$ samples in the same special way as in Fig. 4a. The uniaxial and the hydrostatic pressure data sets show almost the same linear behavior, indicating that in the optimally doped and overdoped region, hydrostatic or uniaxial c-axis pressures give almost identical results.

Indeed the a- and b-axis pressure effects were found to be nearly constant and of opposite sign ($dT_c/dp_a < 0$, $dT_c/dp_b > 0$) for wide ranges of oxygen contents and almost canceling each other in optimally and overdoped samples [43, 59-63]. The sum of a- and b-axis pressure in this doping region results only in a very small pressure effect, which is responsible for the small offset between the hydrostatic and c-axis pressure data in Fig. 4b and can be assigned to $dT_{c,max}/dp$.

From the comparison of uniaxial c-axis and hydrostatic pressure data it can be concluded that in the *optimally* and overdoped region the hydrostatic pressure effect is mainly determined by $dT_c/dp_c$ i.e. by pressure induced charge transfer.

From the uniaxial a- and b-axis pressure data of Kraut et al. [61] it is known that, in contrast to the constant behavior with opposite signs in the optimally and overdoped region, $dT_c/dp_a$ and $dT_c/dp_b$ both have large positive values in the region where the *ortho I / ortho II* transition occurs ($x \approx 6.7$). These large contributions to the hydrostatic pressure effect are responsible for the deviations from the straight line which were found in the underdoped region. The additive effects do not seem to be connected with charge transfer [43, 58]. Calculations of Ludwig et al. [102] showed that for $YBa_2Cu_3O_7$ uniaxial pressure in *a*- or *b*-axis direction does change the relative positions of the apical oxygen atom and the Ba atom between the CuO chain and the $CuO_2$ plane in the same way and therefore does not lead to an effective charge transfer from the CuO chains to the $CuO_2$ planes. As the large positive $dT_c/dp_a$ and $dT_c/dp_b$ values occur just above the onset of the *ortho I / ortho II* transition and decrease strongly for lower hole concentrations [43], again there is evidence that changes in the CuO chain ordering have a strong influence on the variation of $T_c$ under pressure. Pressure induced charge transfer which is surely connected with the application of uniaxial pressure in *c*-axis direction does not seem to be influenced by such a disorder, because $dT_c/dp_c$ is not affected by the *ortho I / ortho II* transition (see [43] and Fig. 4b).

As mentioned in the introduction Ca doping reduces the CuO chain order [88, 91]. Therefore, the decrease of the peak in Fig. 4a with increasing Ca-content can be assigned to this increasing disorder.

To compare our results with the work of Neumeier and Zimmermann [37] and Hikita and Ogasawara [86] we estimated the appropriate $n_h$ values for those samples. We used Eq. (4) and took $T_{c,max}$ of 90.7 K for the La-doped samples of Hikita and Ogasawara. For those samples the oxygen content was almost constant and thus by changing the La-content only, they used a single doping channel, too. In the work of Neumeier and Zimmermann most samples were also doped with Lanthanum and we used the $T_{c,max}$ value of 94.2 K given by Neumeier and Zimmermann. To judge the resulting $n_h$ values we tried to estimate $n_h$ from the chemical doping, too - despite all uncertainties. To consider the distribution of the holes over the CuO chain, the two apex atoms and the two $CuO_2$ planes, we used data of Merz et al. [99] who found each plane to obtain a fraction of approximately 0.22 of each hole in the unit cell of $YBa_2Cu_3O_{6.91}$. The resulting values for the hole content in the $CuO_2$ planes showed a good agreement with the $n_h$ values obtained from Eq. (4), but with a fraction of 0.27 holes dedicated to each $CuO_2$ plane for each hole added by the chemical doping we obtained almost identical results.

By performing our usual method of estimating $n_h$ values from Eq. (4) we fix the $T_c$ maximum to 0.16 holes/$CuO_2$ plane. There are hints that other $n_{h,opt}$ values may be correct for other systems [4, 48, 51, 103, 104] but even if $n_{h,opt}$ is not exactly 0.16 for the Y123 system this would only produce a relative shift of the numbers for the $n_h$ values of our samples. As long as $n_{h,opt}$ is not changed drastically by doping or by pressure application in the Y123 system, the relative $n_h$ values should be correct. From the straight lines in Fig. 4a and b we find that the assumption of an almost constant $n_{h,opt}$ should hold for Y123 with respect to pressure and Ca doping.

In Fig. 5 $dT_c/dp(n_h)$ from our investigation is plotted together with the resulting values from the investigation of Neumeier et al. [37] and Hikita and Ogasawara [86]. There is a good agreement in the overdoped range. In the underdoped range we find the results from Neumeier and Zimmermann and Hikita and Ogasawara to match our results on $Y_{0.89}Ca_{0.11}Ba_2Cu_3O_x$ samples. Considering the average La doping in the underdoped range of the samples of Hikita and Ogasawara and Neumeier and Zimmermann to be approximately 10% La, such an agreement is not surprising. If a disorder in the CuO chain subsystem is really the origin for the suppression of the peak at approximately 0.11 holes per $CuO_2$ plane the amount of this suppression will primary depend on the ability of Ca and La to destroy the CuO chains. As mentioned in the introduction, Ca doped samples are known to have an increased CuO chain disorder and the introduction of La in Y123 causes similar disorder with a tetragonal unit cell for higher La contents [105, 106]. Due to the La content in the underdoped samples of Neumeier and Zimmermann it is understandable that the maximum in $dT_c/dp$ at 0.11 holes per $CuO_2$ plane is suppressed for these samples.

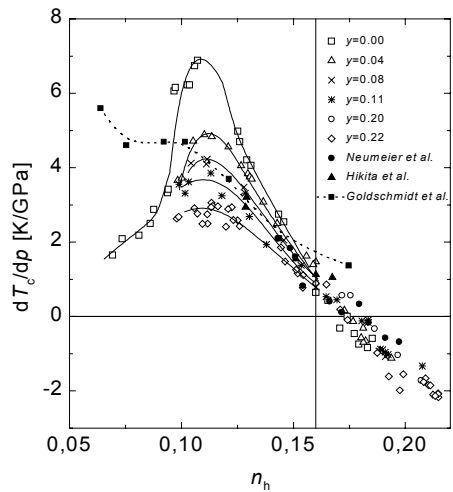

Fig. 5. $dT_c/dp$ of $Y_{1-y}Ca_yBa_2Cu_3O_x$ versus hole concentration $n_h$. For comparison the plot contains values calculated from the data of Neumeier and Zimmermann [37], Hikita and Ogasawara [86] and Goldschmidt et al. [41] as explained in the text.

The values of Goldschmidt et al. shown in Fig. 5 have a higher $dT_{c,max}/dp$ value compared to the doped Y123-samples, which points to the problem of comparing different high-$T_c$ compounds. $dT_{c,max}/dp$ is, to mention but a few, 0.1 K/GPa for $La_{2-z}Sr_zCuO_4$ [50], 1.4 K/GPa for $(Tl_{0.5}Pb_{0.5})Sr_2(Ca_{1-z}Y_z)Cu_2O_7$ [42], 0.015 K/GPa for a Bi 2-layer compound [46], 1.75 K/GPa for a Bi 3-layer compound [51], 2 K/GPa for a Bi 4-layer compound [51], 2.1 K/GPa for the Hg 1-layer compound [48], 1.9 K/GPa for the Hg 2-layer compound [48], and 1.6 K/GPa for the Hg 3-layer compound [48]. Due to this wide spread of $dT_{c,max}/dp$ values the destruction of the CuO chains in Y123 by the charge balanced doping of Goldschmidt et al. is supposed to influence $dT_{c,max}/dp$, too. As a consequence from this argument it may be expected that the $dT_{c,max}/dp$ values for our samples should also be increased with increasing Ca content, but due to the experimental resolution this tendency could not be seen. If we assume $dT_{c,max}/dp$ to contribute to the pressure effect of all samples of Goldschmidt et al. in the same way, we can shift the $dT_c/dp$ values of Goldschmidt et al. in Fig. 5 by -0.9 K/GPa which is the difference of the $dT_{c,max}/dp$ values of the Ca or La doped 123 samples and CaLaBaCuO. Such a uniform shift will produce a curve matching our values for our samples with the highest Ca content for $n_h>0.10$. For lower $n_h$ values the strong $dT_c/dp(n_h)$ decrease of our Ca free sample is not found for CaLaBaCuO. An explanation may be the onset of the ortho II phase for the Ca free Y123. This ordering in the CuO chain sublattice will be suppressed with increasing impurity content and will be completely destroyed for CaLaBaCuO. From this it can be expected that the curves of our Ca doped samples will not match the Ca free values for $n_h<0.11$, but will show an increasing tendency to follow the data of Goldschmidt et al. with larger Ca content. Due to the enormous measuring time necessary for the $T_c$ and $dT_c/dp$ values presented in this paper we could not investigate Ca doped samples with lower $n_h$ values.

Another test of the effect of disordered CuO chains on $dT_c/dp$ is to investigate a sample which is doped by Al, too. In Fig. 6 the $T_c$ values of our samples are shown as a function of $dT_c/dp$ together with the values of the samples of Hikita and Ogasawara [86] and Neumeier and Zimmermann [37]. The curves shown in Fig. 6 are guides to the eye. We obtain curves that are shifted to lower $T_c$ values with higher Ca contents due to the $T_{c,max}$ suppression. In addition the left part of each curve is shifted to the right with increasing Ca content due to the suppression of the $dT_c/dp$ maximum at $n_h \approx 0.11$. The filled circles are data from a $(Y_{0.975}Ca_{0.025})Ba_2(Cu_{2.974}Al_{0.026})O_x$ sample with a Ca content of 2.5% and an additional Al content of 2.6%. Although the total content of dopants is only 5.1% the $T_c$ and $dT_c/dp$ values of these samples match the data of an Al free Y123 sample with approximately 8% Ca. This can be explained by the very strong deterioration effect of Al to the ordering of the CuO chain system [94].

The overdoped samples of Neumeier and Zimmermann probably contain a small amount of Al,

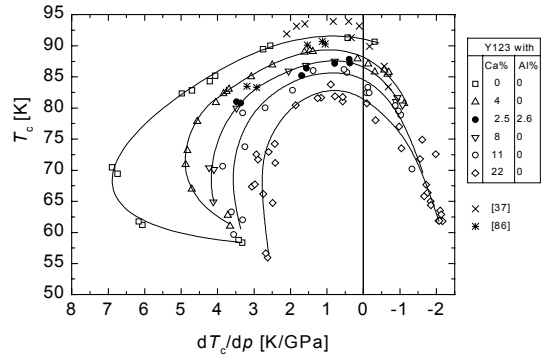

Fig. 6. $T_c$ versus $dT_c/dp$ of $Y_{1-y}Ca_yBa_2Cu_3O_x$ with Ca and Al content as indicated in the table. In addition data points from Neumeier and Zimmermann (x, Ref. [37]) and Hikita and Ogasawara (∗, Ref. [86]) are shown.

too, due to the sample preparation in $Al_2O_3$ crucibles. This may be the reason of the strong $T_c$ suppression for those samples with increasing Ca content. $T_c$ and $dT_c/dp$ of their sample with the highest Ca content of 6% matches our 8% Ca-values, which may be due to the small Al contamination. The La doped samples are difficult to compare in this plot due to the enhanced $T_{c,max}$ value, that may be caused by La on Y sites as mentioned in the introduction. The La doped samples of Hikita and Ogasawara show good agreement with our Ca-doped samples with a corresponding Ca content.

## 4. Conclusions

High-pressure experiments have been performed on $Y_{1-y}Ca_yBa_2Cu_3O_x$ single crystals with $y=0$ to 0.22 to determine $T_c$ and the pressure effect on $T_c$. For each Ca content we investigated single crystals with various oxygen contents $x$ to identify the maximum transition temperature $T_{c,max}(p=0)$ with oxygen as a single doping channel. Due to the additional holes introduced via Ca doping $T_{c,max}$ was found to shift to lower oxygen contents with increasing Ca content. We also found a decrease of $T_{c,max}$ with increasing Ca content.

Using a special plot deduced from the charge transfer model it was shown that for all optimally doped and overdoped $Y_{1-y}Ca_yBa_2Cu_3O_x$ samples a charge transfer of $dn_h/dp \approx 3.7 \cdot 10^{-3}$ $GPa^{-1}$ and a shift of the maximum transition temperature $dT_{c,max}/dp \approx 0.8$ K/GPa can completely describe the experimental results for all Ca concentrations from $y = 0$ to $y = 0.22$. These values are in good agreement with the values of Neumeier and Zimmermann [37] and Hikita and Ogasawara [86].

In the underdoped region, however, additional pressure effects apart from charge transfer lead to large $dT_c/dp$ values. These additional pressure effects are probably caused by the compression of the $a$- and $b$-axes and are strongly pronounced for samples with low Ca contents which have a well developed oxygen order. Deterioration of the oxygen order by additional impurities leads to a decrease of these pressure effects. At low hole concentrations $dT_c/dp$ shows an abrupt decrease to values of approximately 2 K/GPa that may be correlated with the onset of an *ortho* II ordering in the CuO chain system. From the data of Goldschmidt et al. [41] it can be deduced that with inhibition of the *ortho* II ordering the abrupt $dT_c/dp$ decrease vanishes. The results from an Al doped crystal support the importance of intact, but only partially filled CuO chains for the $dT_c/dp$ maximum at $n_h \approx 0.11$, too.

## Acknowledgements

For helpful discussions and the hospitality during his stay at the Department of Physics, Colorado State University, W.H.F. would like to thank Prof. H.D. Hochheimer.